\documentclass{aa}
\usepackage{graphicx}
\usepackage{natbib}
\usepackage{txfonts}

\begin{document}

\title{Comment on \cite{2023A&A...674A.170A}: \\  Large fraction of already known systems reported}
 \author{Zasche,~P.} %~\inst{1}
 \institute{Charles University, Faculty of Mathematics and Physics, Astronomical Institute, V~Hole\v{s}ovi\v{c}k\'ach 2, CZ-180~00, Praha 8, Czech Republic,  \email{zasche@sirrah.troja.mff.cuni.cz} }
 \titlerunning{Comment on \cite{2023A&A...674A.170A}}
 \authorrunning{Zasche P.}
 \date{Received \today; accepted ???}

\abstract{In this work, I report that large fraction of stars detected by \cite[][A\&A, 674,
A170]{2023A&A...674A.170A} and noted in that work as new discoveries are in fact known systems.
This is especially true for the dense bulge fields with large blending of nearby sources. Among
the published 245 stars determined to be doubly eclipsing (i.e. containing two eclipsing signals),
I identified 53 blends. In other words, about a quarter of the systems noted by
\citeauthor{2023A&A...674A.170A} are not actually doubly eclipsing; rather, these are
contaminations of known nearby sources that have already been detected by OGLE. Such a  high
proportion of reported false positives should not be readily ignored and ought to be addressed in
future studies. }

 \maketitle

\section{Introduction}

About one year ago, \cite{2023A&A...674A.170A} published a study on the OGLE-IV catalogue, with an
aim to identify  stars with additional signals besides the dominant one. The authors presented
altogether 292 such new systems, together with their preliminary characteristics and types (i.e.
period and type of variability). According to their statement, they 'determined whether it is a
new variable or just the result of contamination by an already catalogued nearby one' (abstract in
\citealt{2023A&A...674A.170A}). When following up on these results, we came to a very different
conclusion. In addition, those authors noted that they had excluded those that had resulted from
the 'contamination by known OGLE variables from our catalogue'. This amounted to 292 new variable
stars, which the authors listed in their Table B.1. In this work, I present the stars that had
been  missed in this step, which led to a sharp artificial increase in the statistics published in
the cited work.

\begin{table*}[t]%[b]
  \caption{False positives from \cite{2023A&A...674A.170A} together with a blending source identification.}  \label{systems}
  \centering
  \scalebox{0.80}{
  \begin{tabular}{c c c c c c c}%\\[-3mm]
\hline \hline\\[-3mm]
  Target name                 &     RA      &     DE       &   $P_1$   &  $P_2$   &  Blending information     \\
                              &  [J2000.0]  & [J2000.0]    &    [d]    &   [d]    &                           \\   \hline
 \object{OGLE-BLG-ECL-085710} & 17 44 33.76 & -34 06 37.17 & 0.926231  & 0.274207 & blend with close-by OGLE-BLG-ECL-085704 \\
 \object{OGLE-BLG-ECL-162682} & 17 53 34.00 & -31 10 16.82 & 30.509576 & 0.497508 & blend with close-by OGLE-BLG-ECL-162685 \\
 \object{OGLE-BLG-ECL-164230} & 17 53 41.69 & -29 41 38.27 & 1.289267  & 2.990437 & blend with close-by OGLE-BLG-ECL-164216 \\
 \object{OGLE-BLG-ECL-175451} & 17 54 37.17 & -29 25 14.23 & 1.038538  & 0.407096 & blend with close-by OGLE-BLG-ECL-175469 \\
 \object{OGLE-BLG-ECL-175866} & 17 54 39.13 & -29 20 32.21 & 32.155974 & 0.364633 & blend with close-by OGLE-BLG-ECL-175879 \\
 \object{OGLE-BLG-ECL-176073} & 17 54 40.08 & -29 40 49.79 & 0.797012  & 0.351255 & blend with close-by OGLE-BLG-ECL-176113 \\
 \object{OGLE-BLG-ECL-180524} & 17 55 04.07 & -31 00 54.09 & 0.297821  & 0.433165 & blend with close-by OGLE-BLG-ECL-180509 \\
 \object{OGLE-BLG-ECL-183527} & 17 55 19.39 & -29 48 14.56 & 0.336450  & 0.415538 & blend with close-by OGLE-BLG-ECL-183500 \\
 \object{OGLE-BLG-ECL-184951} & 17 55 26.18 & -31 19 32.01 & 0.790206  & 0.389627 & blend with close-by OGLE-BLG-ECL-184961 \\
 \object{OGLE-BLG-ECL-186346} & 17 55 32.94 & -30 00 35.04 & 0.928168  & 0.497266 & blend with close-by OGLE-BLG-ECL-186348 \\
 \object{OGLE-BLG-ECL-195851} & 17 56 20.90 & -28 36 51.08 & 7.321952  & 0.374819 & blend with close-by OGLE-BLG-ECL-195827 \\
 \object{OGLE-BLG-ECL-197037} & 17 56 27.23 & -29 39 37.86 & 1.041253  & 0.336643 & blend with close-by OGLE-BLG-ECL-197030 \\
 \object{OGLE-BLG-ECL-200402} & 17 56 44.16 & -29 40 43.64 & 0.457504  & 0.402577 & blend with close-by OGLE-BLG-ECL-200397 \\
 \object{OGLE-BLG-ECL-202842} & 17 56 56.39 & -30 57 16.88 & 0.565191  & 0.435189 & blend with close-by OGLE-BLG-ECL-202865 \\
 \object{OGLE-BLG-ECL-204851} & 17 57 06.19 & -27 55 51.46 & 0.574790  & 0.430802 & blend with close-by OGLE-BLG-ECL-204861 \\
 \object{OGLE-BLG-ECL-207504} & 17 57 20.89 & -29 25 31.96 & 0.414866  & 0.403268 & blend with close-by OGLE-BLG-ECL-207517 \\
 \object{OGLE-BLG-ECL-207581} & 17 57 21.32 & -29 37 19.21 & 0.369578  & 0.334869 & blend with close-by OGLE-BLG-ECL-207599 \\
 \object{OGLE-BLG-ECL-209728} & 17 57 31.88 & -28 07 54.57 & 0.322124  & 0.336769 & blend with close-by OGLE-BLG-ECL-209712 \\
 \object{OGLE-BLG-ECL-212142} & 17 57 43.88 & -29 26 49.89 & 0.466858  & 0.313969 & blend with close-by OGLE-BLG-ECL-212135 \\
 \object{OGLE-BLG-ECL-213783} & 17 57 52.34 & -29 49 32.35 & 0.415471  & 0.379676 & blend with close-by OGLE-BLG-ECL-213786 \\
 \object{OGLE-BLG-ECL-213786} & 17 57 52.49 & -29 49 33.31 & 0.379677  & 0.415471 & blend with close-by OGLE-BLG-ECL-213783 \\
 \object{OGLE-BLG-ECL-216018} & 17 58 04.54 & -30 51 02.93 & 0.362244  & 0.432300 & blend with close-by OGLE-BLG-ECL-216008 \\
 \object{OGLE-BLG-ECL-216324} & 17 58 06.17 & -28 45 02.19 & 0.643459  & 0.392652 & blend with close-by OGLE-BLG-ECL-216341 \\
 \object{OGLE-BLG-ECL-227744} & 17 59 05.02 & -28 24 39.04 & 1.380442  & 0.419004 & blend with close-by OGLE-BLG-ECL-227750 \\
 \object{OGLE-BLG-ECL-227914} & 17 59 05.80 & -28 27 25.82 & 1.034848  & 0.489439 & blend with close-by OGLE-BLG-ECL-227903 \\
 \object{OGLE-BLG-ECL-229158} & 17 59 12.02 & -29 15 46.01 & 0.620128  & 0.303132 & blend with close-by OGLE-BLG-ECL-229160 \\
 \object{OGLE-BLG-ECL-229495} & 17 59 13.75 & -28 42 17.60 & 0.995846  & 1.288566 & blend with close-by OGLE-BLG-ECL-229478 \\
 \object{OGLE-BLG-ECL-233822} & 17 59 37.95 & -28 22 30.38 & 1.165683  & 0.389816 & blend with close-by OGLE-BLG-ECL-233847 \\
 \object{OGLE-BLG-ECL-235127} & 17 59 44.39 & -29 10 49.55 & 9.048235  & 0.500515 & blend with close-by OGLE-BLG-ECL-235154 \\
 \object{OGLE-BLG-ECL-235154} & 17 59 44.39 & -29 10 49.55 & 0.500514  & 9.047202 & blend with close-by OGLE-BLG-ECL-235127 \\
 \object{OGLE-BLG-ECL-235373} & 17 59 45.46 & -28 35 52.78 & 0.817006  & 0.339319 & blend with close-by OGLE-BLG-ECL-235353 \\
 \object{OGLE-BLG-ECL-236897} & 17 59 53.62 & -28 22 57.27 & 0.461052  & 0.382470 & blend with close-by OGLE-BLG-ECL-236873 \\
 \object{OGLE-BLG-ECL-240175} & 18 00 11.58 & -30 39 20.85 & 0.519637  & 0.575412 & blend with close-by OGLE-BLG-ECL-240177 \\
 \object{OGLE-BLG-ECL-240177} & 18 00 11.58 & -30 39 20.85 & 0.575413  & 0.519634 & blend with close-by OGLE-BLG-ECL-240175 \\
 \object{OGLE-BLG-ECL-245466} & 18 00 39.89 & -28 51 52.88 & 0.685942  & 0.450982 & blend with close-by OGLE-BLG-ECL-245476 \\
 \object{OGLE-BLG-ECL-246036} & 18 00 42.76 & -28 10 14.37 & 0.488428  & 0.718922 & blend with close-by OGLE-BLG-ECL-246041 \\
 \object{OGLE-BLG-ECL-246468} & 18 00 45.10 & -29 03 37.43 & 3.117342  & 0.205621 & blend with close-by OGLE-BLG-ECL-246473 \\
 \object{OGLE-BLG-ECL-251606} & 18 01 11.87 & -28 36 43.44 & 0.334013  & 0.327027 & blend with close-by OGLE-BLG-ECL-251626 \\
 \object{OGLE-BLG-ECL-253194} & 18 01 20.92 & -28 53 57.98 & 0.561837  & 0.324601 & blend with close-by OGLE-BLG-ECL-253185 \\
 \object{OGLE-BLG-ECL-258936} & 18 01 52.69 & -28 34 00.49 & 0.451274  & 0.423425 & blend with close-by OGLE-BLG-ECL-258953 \\
 \object{OGLE-BLG-ECL-259321} & 18 01 54.86 & -27 54 19.18 & 0.446969  & 0.380738 & blend with close-by OGLE-BLG-ECL-259296 \\
 \object{OGLE-BLG-ECL-260224} & 18 01 59.88 & -28 38 10.61 & 0.357231  & 0.197554 & blend with close-by OGLE-BLG-ECL-260240 \\
 \object{OGLE-BLG-ECL-260240} & 18 01 59.88 & -28 38 10.61 & 0.395107  & 0.178616 & blend with close-by OGLE-BLG-ECL-260224 \\
 \object{OGLE-BLG-ECL-266487} & 18 02 34.68 & -27 47 24.50 & 0.760567  & 0.360594 & blend with close-by OGLE-BLG-ECL-266498 \\
 \object{OGLE-BLG-ECL-269995} & 18 02 54.52 & -26 46 02.65 & 0.656073  & 0.511003 & blend with close-by OGLE-BLG-ECL-269991 \\
 \object{OGLE-BLG-ECL-270233} & 18 02 55.86 & -27 20 34.68 & 2.200611  & 0.431809 & blend with close-by OGLE-BLG-ECL-270203 \\
 \object{OGLE-BLG-ECL-272656} & 18 03 09.46 & -28 46 38.54 & 3.385598  & 0.362125 & blend with close-by OGLE-BLG-ECL-272668 \\
 \object{OGLE-BLG-ECL-274965} & 18 03 22.32 & -28 38 37.58 & 1.341903  & 0.387542 & blend with close-by OGLE-BLG-ECL-274953 \\
 \object{OGLE-BLG-ECL-279001} & 18 03 44.10 & -28 45 39.15 & 0.815092  & 0.241045 & blend with close-by OGLE-BLG-ECL-279020 \\
 \object{OGLE-BLG-ECL-280921} & 18 03 54.55 & -28 45 05.97 & 8.075665  & 0.356790 & blend with close-by OGLE-BLG-ECL-280936 \\
 \object{OGLE-BLG-ECL-285403} & 18 04 19.59 & -27 57 03.77 & 1.077844  & 0.157675 & blend with close-by OGLE-BLG-ECL-285426 \\
 \object{OGLE-BLG-ECL-286273} & 18 04 24.13 & -29 11 50.39 & 0.256666  & 0.408718 & blend with close-by OGLE-BLG-ECL-286294 \\
 \object{OGLE-BLG-ECL-293405} & 18 05 03.12 & -29 09 14.25 & 0.284271  & 0.423757 & blend with close-by OGLE-BLG-ECL-293399 \\
 \hline
\end{tabular}} %\\
 \end{table*}

\section{Known nearby variables}

For the detection of known variables in the vicinity of the particular target, we also used the
same source as \cite{2023A&A...674A.170A}, namely, is the OGLE-IV catalogue from
\cite{2016AcA....66..405S}. The area scanned for detections was set to 10$^{\prime\prime}$ around
the target, where I tried to identify the source of the signal from \cite{2016AcA....66..405S}
with the same period as given by \cite{2023A&A...674A.170A}.

The serious blending problem is usually related to the photometric data obtained on large pixels
with poor angular resolution (e.g. TESS with its 21$^{\prime\prime}$/px, see
\citealt{2015JATIS...1a4003R}). However, for the OGLE-IV data, this is problematic only in very
dense stellar fields in the galactic bulge. In controlling all the presented 245 new systems with
two eclipsing periods by \cite{2023A&A...674A.170A}, there were  53 nearby sources of this
additional signal identified. Hence, the 53 false positives presented by
\cite{2023A&A...674A.170A} represents almost a quarter of all their reported systems. These stars
are given in Table \ref{systems}, along with the true identification of the source of additional
variation.

 \begin{figure}
  \centering
  \begin{picture}(280,160)% width and height of the picture
  \put(0,82){
   \includegraphics[width=0.24\textwidth]{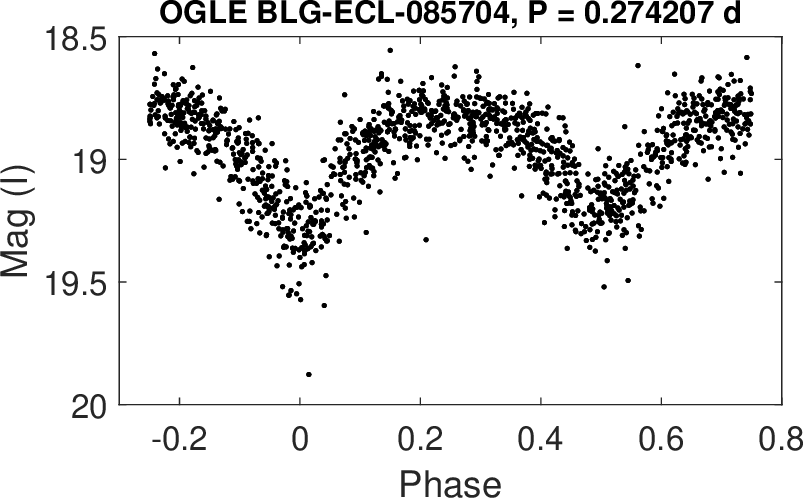}}
   \put(0,0){
   \includegraphics[width=0.24\textwidth]{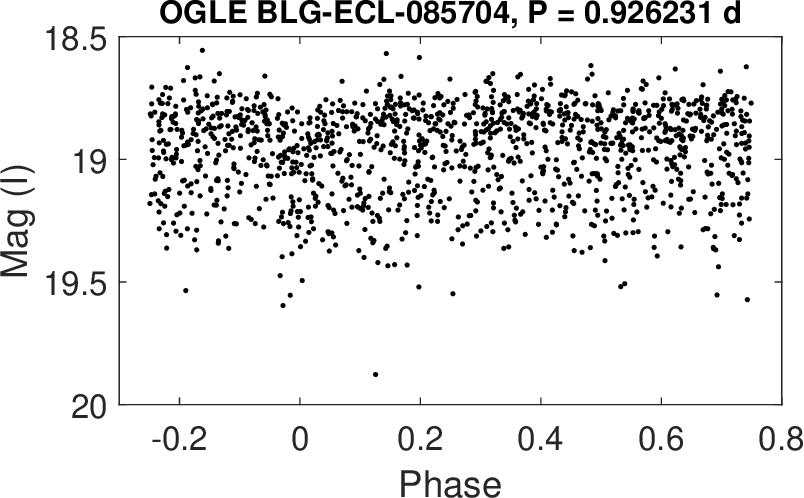}}
   \put(128,82){
   \includegraphics[width=0.24\textwidth]{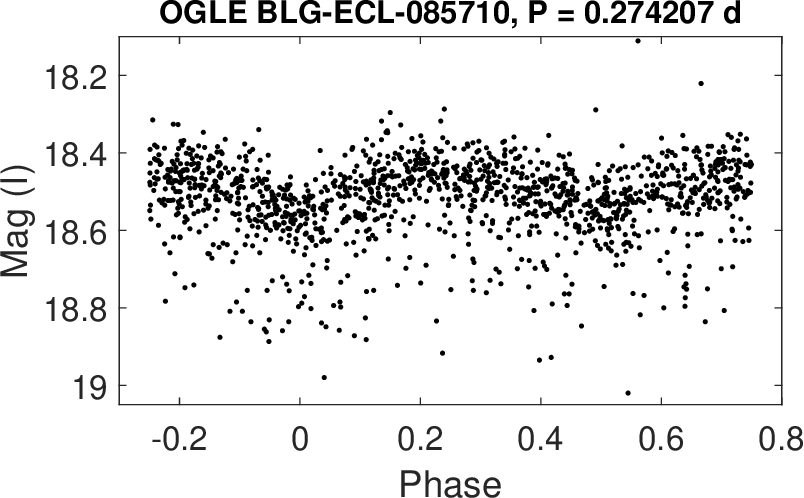}}
   \put(128,0){
   \includegraphics[width=0.24\textwidth]{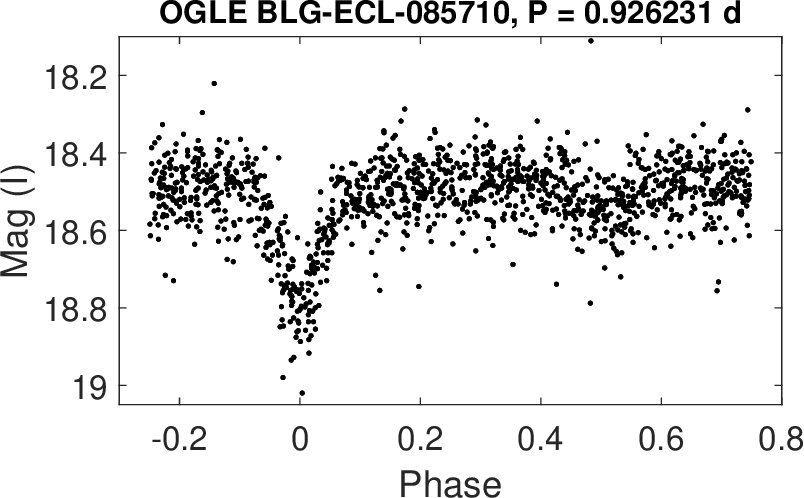}}
  \end{picture}
  \caption{Typical example of two close stars OGLE-BLG-ECL-85704 and OGLE-BLG-ECL-85710 (separated about 1.07$^{\prime\prime}$) with two different periods. Each of the periods clearly belongs to one particular target and the residual signal is only due to light contamination.}
  \label{FigLC}
 \end{figure}

One typical example of such a false positive is shown in Figure 1, drawn from the first system in
our Table 1. We can clearly see the appropriate period for the particular star. Both  signals are
present on both stars thanks to their close angular distance (about 1.07$^{\prime\prime}$ only),
but identifying   which star the period belongs to is straightforward.

As an additional cross-check, we also tried to identify similar periods (or their double and/or
half values) from other photometric data. We used the old ASAS-3 database
\citep{1998AcA....48...35P,2002AcA....52..397P}, searching in the  2$^{\prime}$ radius around a
star, as well as the ASAS-SN database  \citep{2017PASP..129j4502K,2018MNRAS.477.3145J}, also
within the same radius. Finally, we used \textit{Gaia} DR3 \citep{2023A&A...674A...1G} and stars
in the 70$^{\prime\prime}$ radius were scanned. As a result, no other positive detection of both
eclipsing periods in these databases was found. From the remaining 192 systems for 23 stars, their
more prominent periods were also identified in the \textit{Gaia} DR3 variability tables.

\section{Conclusions}  \label{discussion}

The blending problem of light contamination from nearby sources is less obvious for the OGLE-IV
data (angular resolution of about 0.417$^{\prime\prime}$/px) than for TESS data. However, in dense
stellar fields, this can pose a serious problem to classification procedures. Therefore, it is
important to make an effort to identify false positives such as those presented in the paper by
\cite{2023A&A...674A.170A}, as a necessary step in any such analysis.

\begin{acknowledgements}
We do thank the OGLE team for making all of the observations easily public available. This
research has made use of the SIMBAD and VIZIER databases, operated at CDS, Strasbourg, France and
of NASA Astrophysics Data System Bibliographic Services. This work has made use of data from the
European Space Agency (ESA) mission {\it Gaia} (\url{https://www.cosmos.esa.int/gaia}), processed
by the {\it Gaia} Data Processing and Analysis Consortium (DPAC,
\url{https://www.cosmos.esa.int/web/gaia/dpac/consortium}). Funding for the DPAC has been provided
by national institutions, in particular the institutions participating in the {\it Gaia}
Multilateral Agreement.
\end{acknowledgements}


\begin{thebibliography}{}
  \bibitem[{\'A}d{\'a}m et al.(2023)]{2023A&A...674A.170A} {\'A}d{\'a}m, R.~Z., Hajdu, T., B{\'o}di, A., et al.\ 2023, \aap, 674, A170%. doi:10.1051/0004-6361/202346006
  \bibitem[Gaia Collaboration et al.(2023)]{2023A&A...674A...1G} Gaia Collaboration, Vallenari, A., Brown, A.~G.~A., et al.\ 2023, \aap, 674, A1
  \bibitem[Jayasinghe et al.(2018)]{2018MNRAS.477.3145J} Jayasinghe, T., Kochanek, C.~S., Stanek, K.~Z., et al.\ 2018, \mnras, 477, 3145%. doi:10.1093/mnras/sty838
  \bibitem[Kochanek et al.(2017)]{2017PASP..129j4502K} Kochanek, C.~S., Shappee, B.~J., Stanek, K.~Z., et al.\ 2017, \pasp, 129, 104502
  \bibitem[Pojmanski(1998)]{1998AcA....48...35P} Pojmanski, G.\ 1998, \actaa, 48, 35%. doi:10.48550/arXiv.astro-ph/9802330
  \bibitem[Pojmanski(2002)]{2002AcA....52..397P} Pojmanski, G.\ 2002, \actaa, 52, 397%. doi:10.48550/arXiv.astro-ph/0210283
  \bibitem[Ricker et al.(2015)]{2015JATIS...1a4003R} Ricker, G.~R., Winn, J.~N., Vanderspek, R., et al.\ 2015, JATIS, 1, 014003%. doi:10.1117/1.JATIS.1.1.014003
  \bibitem[Soszy{\'n}ski et al.(2016)]{2016AcA....66..405S} Soszy{\'n}ski, I., Pawlak, M., Pietrukowicz, P., et al.\ 2016, \actaa, 66, 405%. doi:10.48550/arXiv.1701.03105
 \end{thebibliography}
\end{document}